# Spin-resolved imaging of atomic-scale helimagnetism in monolayer NiI$_2$

Mao-Peng Miao[1], Nanshu Liu[2,3], Wen-Hao Zhang[1], Dao-Bo Wang[1], Wei Ji[2,3] and Ying-Shuang Fu[1,4,5*]

1. School of Physics and Wuhan National High Magnetic Field Center, Huazhong University of Science and Technology, Wuhan 430074, China

2. Beijing Key Laboratory of Optoelectronic Functional Materials & Micro-Nano Devices, Department of Physics, Renmin University of China, Beijing 100872, China

3. Key Laboratory of Quantum State Construction and Manipulation (Ministry of Education), Renmin University of China, Beijing, 100872, China

4. Hubei Key Laboratory of Gravitation and Quantum Physics, Huazhong University of Science and Technology, Wuhan 430074, China

5. Wuhan Institute of Quantum Technology, Wuhan 430206, China

*Email: yfu@hust.edu.cn

**Abstract:**

Identifying intrinsic noncollinear magnetic order in monolayer van der Waals (vdW) crystals is highly desirable for understanding the delicate magnetic interactions at reduced spatial constraints and miniaturized spintronic applications, but remains elusive in experiments. Here, we achieved spin-resolved imaging of helimagnetism at atomic scale in monolayer NiI$_2$ crystals, that were grown on graphene-covered SiC(0001) substrate, using spin-polarized scanning tunneling microscopy. Our experiments identify the existence of a spin spiral state with canted plane in monolayer NiI$_2$. The spin modulation Q vector of the spin spiral is determined as (0.2203, 0, 0), which is different from its bulk value or its **in-plane projection, but agrees well with our first principles calculations. The spin spiral surprisingly indicates collective spin switching behavior under magnetic field, whose origin is ascribed to the incommensurability between the spin spiral and the crystal lattice. Our work unambiguously identifies the helimagnetic state in monolayer $NiI_2$, paving the way for illuminating its expected type-II multiferroic order and developing spintronic devices based on vdW magnets.**

Investigating intrinsic magnetic order in two-dimensional (2D) systems provides insights into the fundamental understanding of magnetic interactions and fluctuations influenced by the reduced spatial constraints, and also drives technological advances in building miniaturised spintronic devices. According to the Mermin-Wagner theorem [1], long-range order of the magnetic moments cannot exist at finite temperatures in low dimensions due to low energy excitations agitated via thermal fluctuations. Nevertheless, the fluctuations can be suppressed with magnetic anisotropy, which gaps out low energy excitations and renders the long range 2D magnetic order survive. The recent discovery of 2D ferromagnetism in ultrathin van der Waals crystals of $CrI_3$ [2] and $Cr_2Ge_2Te_6$ [2] has triggered tremendous interests in searching magnetism in other 2D vdW materials, revealing a wealth of emergent phenomena that are not observed in their 3D counterparts [4,5]. Among them, the stability and nature of the 2D magnetic order indicates versatile tenability via external perturbations, such as interlayer coupling [6,7], electrostatic gating [8,9], and the presence of defects [10,11], etc.

While collinear ferromagnetic and antiferromagnetic orders have been discovered in numerous 2D vdW magnets, the noncollinear magnetic order is rarely explored experimentally [12-14], presumably because its existence demands intricate magnetic interactions and lack of appropriate experimental probes. Noncollinear magnetism arises from the delicate balance of Heisenburg exchange interactions between nearest neighbor and higher neighbors that leads to spin frustration or from the involvement of competing antisymmetric Dzyaloshinskii–Moriya interaction due to inversion symmetric breaking and spin-orbital coupling [15,16]. In addition, the characterization of noncollinear magnetism is also challenging because most of the magnetic sensitive techniques such as neutron scattering cannot be applied to monolayer films. Ensemble averaged optical probes such as magnetic circular dichroism and second harmonic generation (SHG) have been used in characterizing noncollinear magnetism [12-14]. But they rely on the comparison with theoretical modeling and cannot determine specific Q vectors that characterize the propagating direction and period of the noncollinear magnetic order. While recent quantum magnetometry based on nitrogen-vacancy has witnessed its success in charactering the moiré magnetism in twisted $CrI_3$ bilayers [17], its spatial resolution is insufficient to resolve the theoretically anticipated noncollinear magnetism [18]. Spin-polarized scanning tunneling microscopy (SPSTM) can directly identify the magnetic order with atomic-scale spin resolution based on its record for spin-dependent electron tunneling currents [18], rendering it an ideal probe to characterize magnetic orders virtually for all different types.

Helimagnetism is a typical noncollinear magnetic order [20], in which constitute local spins gradually rotate on its spin spiral plane along the array of magnetic ions, causing the trajectory of the spin vectors to form a spiral. Our system of choice is monolayer $NiI_2$, because its bulk form supports helimagnetic order below 59.5 K that may survive down to monolayer [21-23]. Moreover, monolayer $NiI_2$ is expected to be a type-II multiferroic candidate, whose ferroelectric polarization is magnetism-driven and is oriented orthogonal to the Q vector of the spin spiral. While recent SHG studies suggest the existence of multiferroelectric state in monolayer $NiI_2$ [14], interpretations of the experimental results are in debate [24], making the precise determination of its spin state crucial to resolve the controversy.

Here, using SPSTM, we investigate the spin texture state in monolayer $NiI_2$ at atomic scale. Our experiments unveil that monolayer $NiI_2$ has a spin spiral state whose rotation plane is canted relative to the film surface. The Q vector of the spin spiral is determined as (0.2203, 0, 0), corresponding to a period of 4.54 times the lattice constant of the Ni ions, $a_0$, which is different to its bulk value (0, 0.1384, 1.457) or its in-plane projection, but is in nice agreement with our calculated spin spiral period of $4.5a_0$ with density functional theory (DFT). Moreover, the spin moments of the monolayer $NiI_2$ can be collectively switched between their bistable configurations by an out-of-plane magnetic field with a threshold value of about 2 T. Such stunning field-induced switching behavior is associated to the incommensurability between the spin spiral and the crystal lattice.

The experiments were performed with a custom-made Unisoku STM (1500) system at 2 K [7]. High quality monolayer $NiI_2$ films were grown by evaporating $NiI_2$ powders with molecular beam epitaxy on graphene-covered SiC(0001) substrate. The theoretical calculation is carried out with DFT plus *U*. Detailed descriptions of the experiments and the calculations are depicted in Supplementary Information.

Bulk $NiI_2$ is a vdW crystal that belongs to trigonal $\bar{R}3m$ space group ($CdCl_2$ prototype) at room temperature. Its each vdW layer consists of a triangular lattice of Ni layer sandwiched between two I layers, with each Ni cation ($3d^8$, S=1) octahedrally coordinated by six I anions, forming a 1-T structure [Fig. 1(a)]. With decreasing the temperature, bulk $NiI_2$ undergoes two successive magnetic phase transitions to an interlayer AFM state at 75 K and a proper-screw helimagnetic state with a Q vector of (0, 0.1384, 1.457) at 59.5 K [21-23]. While the helimagnetic state in bulk $NiI_2$ has been well-established, the magnetic ground state of its monolayer form remains an open issue. Various magnetic configurations, ranging from ferromagnetism [25-27], antiferromagnetism [28] to different types of helimagnetism [29-30], have been predicted in DFT calculations.

We first characterize the morphology of the monolayer $NiI_2$ films with conventional W tips. $NiI_2$ films preferentially nucleate at step edges of the graphene substrate. Its typical STM topographic image displays atomically flat surfaces [Fig. 1(b)], whose apparent height is measured as 750 pm at a sample bias of 2 V (Supplementary Fig. S1), conforming to the its expected monolayer height. An atomic-resolution STM image of the monolayer $NiI_2$ film shows its in-plane lattice constant of I atoms is about 388 pm [Fig. 1(c)], which is consistent with the theoretical value of 391 pm [31]. The monolayer $NiI_2$ films exhibit multiple stacking angles relative to the graphene substrate (Supplementary Fig. S2). This implies the interaction between the film and substrate is weak, ensuring the intrinsic magnetic properties of monolayer $NiI_2$.

The monolayer $NiI_2$ surfaces contain a considerable density of dark spots [Fig. 1(d)]. Those dark spots vanish upon imaging the same surface region at low sample biases (Supplementary Fig. S3), revealing intact atomic lattice of top layer I. Those spots can be controllably manipulated with the STM tip (Supplementary Fig. S3), indicating they are polarons, instead of crystal defects. We have discovered and manipulated similar polarons in $CoCl_2$ and $FeCl_2$ films [32]. Tunneling spectrum acquired at the polaron-free region of the surface features a large insulating gap of 2.13 eV, where the conduction band minimum and valence band maxim are located at 0.25 and -1.78 eV, respectively.

Having characterized the morphology and electronic states of monolayer $NiI_2$, we investigate its magnetic spin state with SPSTM. For that, a Fe-coated tip is firstly used. Fe-coated tip possesses in-plane spin-sensitivity due to its shape anisotropy and has the virtue of large spin polarization. More importantly, its magnetization direction can be aligned along external magnetic fields [33-34]. Since the prominent polaron contrast may obscure the weaker spin contrast, we first "clean" polarons out of the selected region of interest via tip manipulation. Fig. 2(a) shows an SPSTM image of such a cleaned surface region at +650 mV. The image displays clear stripe patterns. Upon application of a -1 T field, the stripe pattern shifts drastically relative to a defect marker [Fig. 2(b), red circle]. The marker is from a real crystal defect and cannot be manipulated by the tip. Such stripe patterns are not observable with conventional nonmagnetic W tips. This, in conjunction with its shifts under magnetic field [Fig. 2(a) and 2(b)], unambiguously demonstrates the stripe pattern is a spin contrast.

The strip pattern could be a spin contrast from an AFM state or a spin spiral. To differentiate those two states, we have systematically investigated the response of the stripe pattern to different magnetic fields, where the detailed SPSTM images and conductance mappings are in Supplementary Figs. S4 and S5. As shown in Figs. 2(c,d), the stripe pattern progressively shifts relative to the defect marker with increasing magnetic field and reaches a saturation at 0.5 T, which field is typical for fully aligning the Fe tip to out-of-plane [33-34]. The shifting directions and saturation fields are opposite for different field polarities. This demonstrates the stripe pattern is from a spin spiral, instead of an AFM state. Because the AFM spin contrast does not shift continuously with the different magnetic fields. As will be shown later, the spin spiral keeps its spin magnetizations unchanged within the magnetic field range of [-2, +2] T (Supplementary Fig. S6). Thus, the stripe pattern shift is all caused by the field-induced rotation of the tip magnetization.

Spatial period of the stripes is measured as 1.76 nm, corresponding to $4.54a_0$. As is determined from a high-resolution SPSTM image and its Fourier transformation [Figs. 2(e,f)], the stripe pattern are along a 30° direction off the I lattice. Consequently, the Q vector of the spin spiral is determined as (0.2203, 0, 0).

Apart from the Q vector, another essential feature that characterizes the spin spiral is the spin-spiral plane, on which the spins rotate [20]. For SPSTM measurements, the spin-polarized tunneling current ($I_{SP}$) is sensitive to the relative angle ($\varphi$) between the magnetization of the tip ($P_{tip}$) and the sample ($P_{sample}$), describing as $I_{SP} \propto P_{tip} \cdot P_{sample}$

∝cosφ [19]. Namely, the spin contrast is brightest (darkest) at φ = 0°(180°), and disappears at φ = 90°. As such, we applied a small field to tilt the tip magnetization to determine the spin-spiral plane, because the spin contrast would vanish upon the tip magnetization is adjusted perpendicular to the spin-spiral plane. Fig. 2(g) shows such a SPSTM image in derivative mode, displaying two neighboring stripe domains with an intersect angle of 120° at 0 T. The raw SPSTM image is in Supplementary Fig. S7. Upon application of a 0.1 T field, stripe pattern of the upper domain vanishes, but is kept visible at the lower domain [Fig. 2(h)]. This demonstrates the spin-spiral plane of the upper domain is orthogonal to the tip magnetization, which is canted. Thus, the spin spiral plane should also be canted relative to the film surface, as is depicted in Fig. 2(i).

Since the Fe-coated tip still has a finite stray field that may potentially perturb the spin spiral state, we have further performed SPSTM measurements with AFM Cr-coated tips, whose stray field is negligibly small. Thus, the Cr-tip magnetization, despite usually being canted, is invariant against magnetic field, rendering it easily disentangled from the spin state changing from the sample. Figs. 3(a-d) show series of SPSTM images taken with the Cr tip on the same $NiI_2$ region under four typically selected magnetic fields, showing clearly resolved stripe patterns relative to a defect marker. In accordance with the stripe pattern, the spin polarized spectra at different locations of the pattern also display obvious differences, revealing spin asymmetry as a function of energy [Fig. 3(e)]. As expected for the Cr tip magnetization, the stripe pattern is robust against the magnetic field between -2 T and +2 T, demonstrating the spin spiral is unchanged (Supplementary Fig. S6). However, the stripe pattern abruptly reverses its contrast upon the magnetic field exceeds 2.1 T, and keeps the same afterwards up to our maximum applied field of 4 T. Similar contrast reversal occurs at -2.0 T, resulting a hysteresis loop of the contrast [Fig. 3(f)]. Here, we define the sample state as "1 (0)" under a +4 T (-4 T) magnetic field. Such spin contrast switching is reproduced with different Cr tips and monolayer $NiI_2$ films, showing similar threshold field values. And, the SPSTM experiment with Fe-coated W tips (supplementary Fig. S8) yields consistent results, indicating the switching behavior is an intrinsic property to the spin spiral.

Previous studies on spin spiral states indicate their local magnetizations, instead of switching, are titled towards the magnetic field direction, resulting changes in their Q vectors [35-38]. The surprising magnetization reversal of our spin spiral demonstrates it has residual magnetic moment and its neighboring spin moments are strongly coupled. We conjecture that residual moment of the spin spiral is rooted in its the incommensurability with the crystal lattice. Due to the incommensurability, the spin-up component of the SDW state, that is integrated over a mostly matched lattice period, is larger than its spin-down counterpart, giving rise to the residual moment. Such spin imbalance eventually disappears in integer number of crystal periods that reaches commensurability, which corresponds to $454a_0$ ($= 176$ nm) in the current system. However, such stringent requirements cannot be satisfied in actual $NiI_2$ domains of finite size, rendering the residual moments survive and the magnetization reversal feasible.

To substantiate the experimental observations, we have conducted DFT

calculations on a freestanding NiI$_2$ monolayer. As shown in Fig. 4(a), we found a new spiral order of Q=(0.22, 0, 0), which propagates along the *x* direction across each 9 × √3 supercell with magnetic moments rotating by an angle of 4π in the Ni-I plane [upper panel in Fig. 4(a)]. A stripe characteristic of magnetization density along the *x* direction is observed with a width of about 1.77 nm (lower panel in Fig. 4a), well consistent with the experimental value of 1.76 nm. This spiral order is almost degenerate in energy with the spin spiral configuration (Q=0.25, 0, 0) found in our previous theoretical work [39]. Such a non-collinear magnetic ground state of NiI$_2$ monolayer is mainly due to a competition between FM $J_1$ and AFM $J_3$ with a |$J_1/J_3$| value of 1.26 [39], which well fits a criterion that non-collinear spin spiral order will be favored if | $J_1/J_3$| < 4 [40]. $J_1$, $J_2$ and $J_3$ are the nearest, second nearest and third nearest neighbor isotropic exchange interactions, respectively [see Fig. 1(a)]. The spiral plane [Fig. 4(b)] is canted to the Ni-I plane, conforming to the experimental observation. The canted spiral plane is caused by a significant Kitaev interaction in NiI$_2$. A recent theoretical study also suggests that the Kitaev interaction plays a key role in NiI$_2$ and is proved to impose anisotropy on coplanar spin texture [29]. The magnetic moments along the three directions of spin spiral configuration (Q=0.22, 0, 0) in every 4.5 × √3 supercell are incommensurate [Fig. 4(c)], giving a net magnetic moment of 0.18 $\mu_B$ per 4.5 × √3 supercell, which guarantees the response of the different stripe patterns under magnetic fields applied in the experiment [Fig. 3].

In summary, we have identified the existence of a spin spiral state with canted plane in monolayer NiI$_2$, demonstrating the dominant role of Kitaev interaction. To our knowledge, this is the first observation of atomic-scale noncollinear magnetic order at monolayer limit. The determined Q vector of the monolayer spin spiral is different from the bulk value or its in-plane projection, but agrees nicely with our DFT calculations. Our work clarifies the controversies over the magnetic configuration of monolayer $NiI_2$, which is crucible for understanding its anticipated type-II multiferroic state. We have also attempted unveiling the expected ferroelectricity via measuring the local band bending adjacent to the film edges and domain boundaries (Supplementary Fig. S9). However, no observable ferroelectric polarization was observed, presumably because its value is 3~4 orders of magnitude smaller than typical type-I ferroelectric materials. [24,41-42]. The spin spiral surprisingly indicates tunable bistability with magnetic field, whose origin is ascribed to the spiral incommensurability. This offers a distinct route to manipulate the spin spiral state from previous studies, and expects to be generalized to other incommensurate spin spiral materials. In addition to the studies of fundamental physics in helimagnetism such as magnon excitations, this system also provides a platform for studying spin states of single polarons at 2D limit, as well as the development of spintronic devices based on vdW magnets.

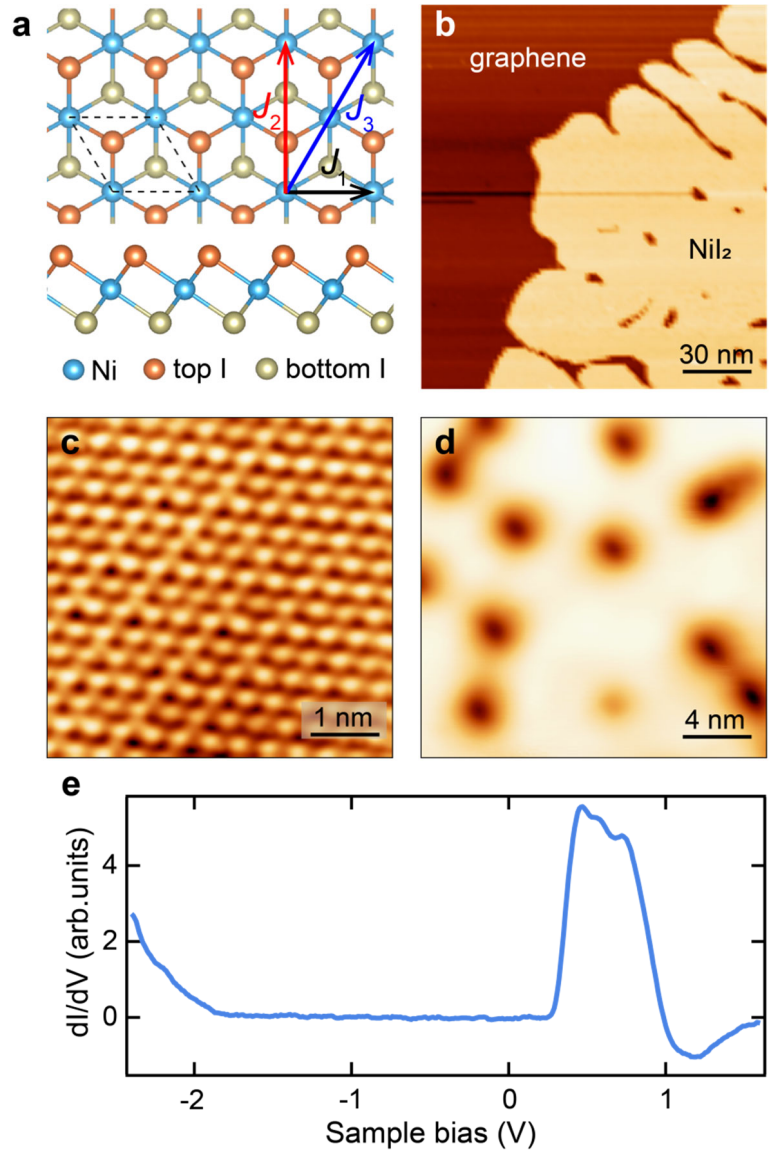

**Figure 1. Morphology and electronic structure of monolayer NiI₂ measured with a conventional W tip. a,** Top view (up) and side view (down) of the crystal structure of monolayer NiI₂. $J_1$, $J_2$ and $J_3$ are the isotropic Heisenberg exchange parameters. **b,** Large-scale STM topographic image of NiI₂ film ($V_s$ = 2 V, $I_t$ = 5 pA). **c,** Atomic resolution of the STM image ($V_s$ = 0.1 V, $I_t$ = 20 pA) of monolayer NiI₂. **d,** Spin-averaged STM image ($V_s$ = 0.65 V, $I_t$ = 10 pA) of NiI₂. **e,** Tunneling spectra ($V_s$ = 1.6 V, $I_t$ = 100 pA, $V_{mod}$ = 30 mV) measured on monolayer NiI₂ film.

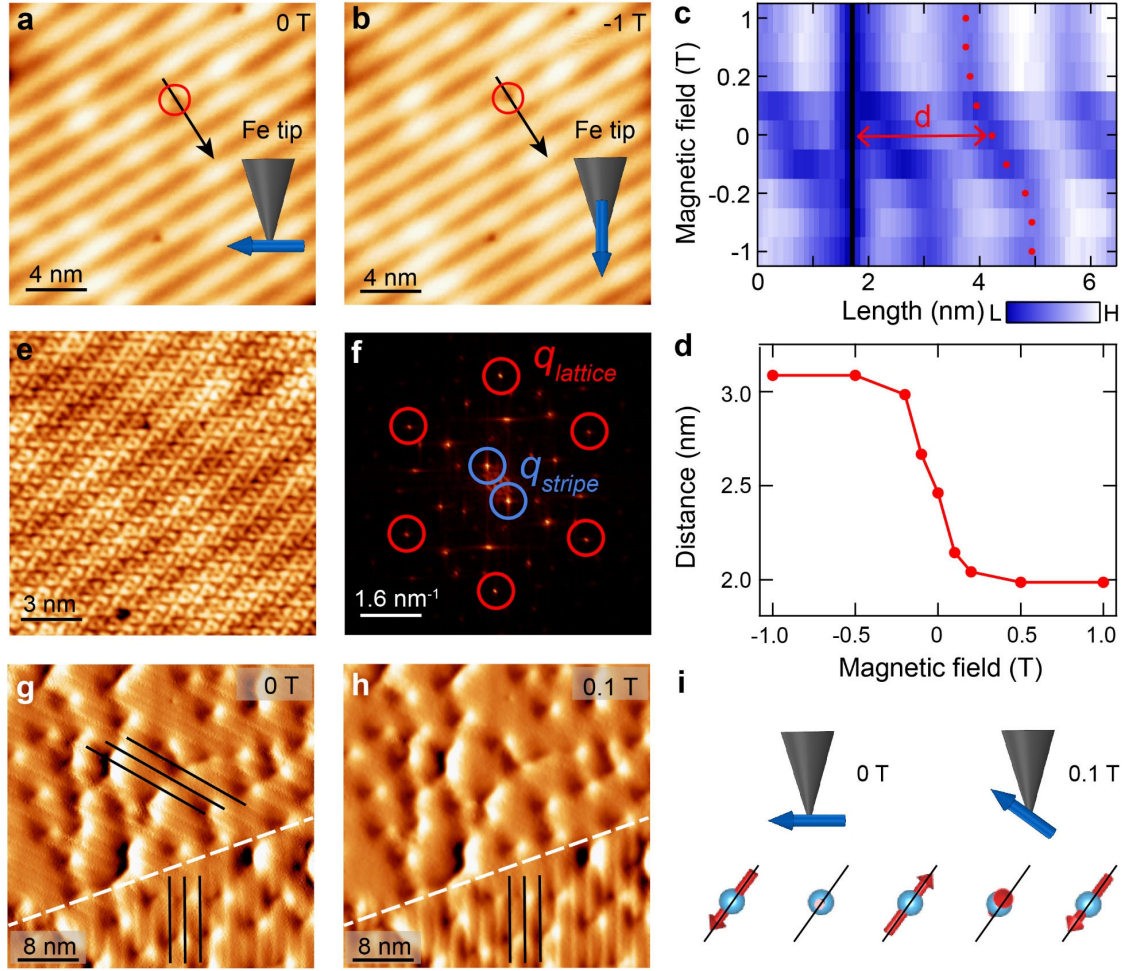

**Figure 2. Spin mapping of the spin spiral texture in monolayer NiI$_2$ with an Fe tip.**

**a,b,** SPSTM images ($V_s$ = 0.65 V, $I_t$ = 10 pA) of monolayer NiI$_2$ under different magnetic fields. An I vacancy defect is marked with red circles. The Fe-tip magnetization under different magnetic fields is depicted with blue arrows. **c,** 2D plot of line profiles taken across the defect marker (black lines in **a** and **b**) under different magnetic fields. Red dots mark locations a selective dark stripe, whose distance, d, relative to the marker is marked with a red arrow segment. **d,** Statistics of the distance between the marker and the selected dark stripe under different magnetic fields. **e,** High-resolution of SPSTM image ($V_s$ = 0.2 V, $I_t$ = 10 pA) of monolayer NiI$_2$, showing coexistence of the atomic lattice and stripe pattern **f,** FFT image of (**e**). The diffraction spots for the lattice $q_{lattice}$ and the spin stripe $q_{stripe}$ are marked by red and blue circles, respectively. **g, h,** SPSTM images ($V_s$ = 0.65 V, $I_t$ = 10 pA) in derivative mode showing two neighboring stripe domains in monolayer NiI$_2$ at 0 T and 0.1 T, respectively. The stripe patterns are selectively marked with black lines. White dashed lines represent the domain boundary. Note that polarons are not "cleaned" in this region. **i,** Schematics of the SPSTM measurement configurations under different magnetic fields. The Fe-tip magnetization P$_{tip}$ and the sample magnetization P$_{sample}$ are represented with blue and red arrows, respectively. The parallel black segments depict the spin spiral plane.

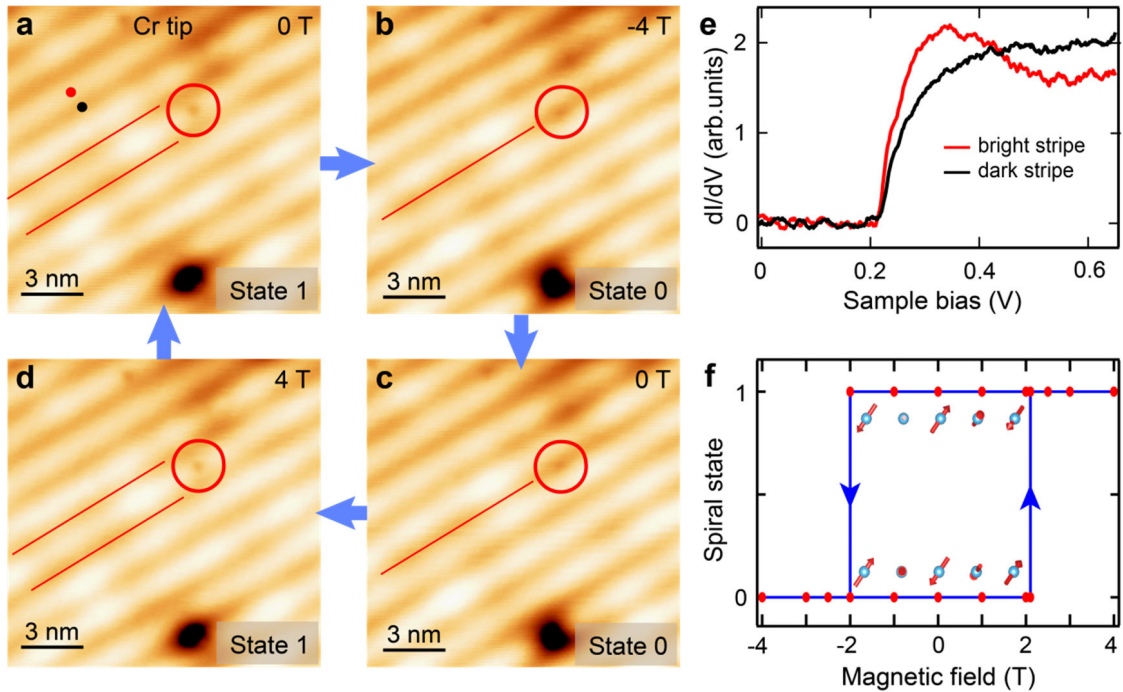

**Figure 3. Bistability of the spin spiral state in monolayer NiI$_2$ measured with a Cr tip. a-d,** SPSTM images ($V_s$ = 0.65 V, $I_t$ = 10 pA) of monolayer NiI$_2$ taken with a Cr tip under different magnetic fields. An I vacancy is marked with red circles as a marker. The spin spiral state is defined as "state 1" ("state 0") for the case of the defect marker

residing on the bright (dark) stripes. The red lines selectively mark representative dark stripes. **e,** Spin-polarized tunneling spectra measured on the bright (red) and the dark stripe (black) of monolayer $NiI_2$ film, whose spectroscopic locations are marked in (**a**). **f,** Statistics of the spin spiral state under different magnetic fields. The opposite local moments between the spin spiral "state 1" and "state 0" are depicted in the inset.

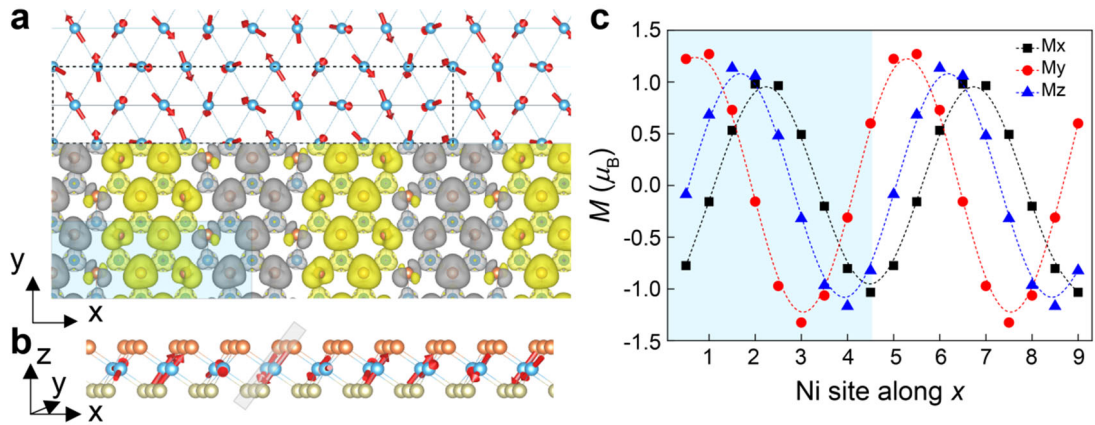

**Figure 4. DFT-calculated spin spiral of monolayer $NiI_2$. a,** Top view of spin spiral order (Q=0.22, 0, 0). The $9 \times \sqrt{3}$ supercell is labeled by a dashed black rectangular. The lower panel is magnetization density along the *x* direction with an isosurface value of $5 \times 10^{-4}$ $e$/Bohr$^3$. The yellow and grey contours stand for positive and negative moments, respectively. The $4.5 \times \sqrt{3}$ supercell is also labeled by a black rectangular. **b,** Perspective view of the spin spiral order (Q=0.22, 0, 0). The spin spiral plane is illustrated by a grey parallelogram. **c,** Magnetic moments projections in the three directions for the Ni atoms labeled in (**a**). A $4.5 \times \sqrt{3}$ supercell is shown by a shadow rectangular.

**Methods**

**Molecular-beam epitaxy (MBE) growth.** The graphene substrate was obtained by high-temperature pyrolysis of SiC(0001) substrate with cycles of vacuum flashing treatment, the details of which are provided in [43]. The 1T-$NiI_2$ films were grown on the SiC supported bilayer graphene by evaporating $NiI_2$ powders (purity, 99.99%) with molecular beam epitaxy. The substrate was kept at 373 ~ 423 K to deposit $NiI_2$.

**STM measurements.** The measurements were performed in a custom-made Unisoku STM (1500) system [7] mainly at 2 K unless described exclusively. The spin-averaged STM data were measured with an electrochemically etched W tip. The SPSTM data were taken with Cr-coated or Fe-coated W tips. All the W tip were flashed to ~2000 K to remove oxides. While the Cr tip was prepared by coating ~10 layers of Cr (purity: 99.995%) on a flashed W tip, the Fe tip was prepared by coating ~30 layers of Fe on a flashed W tip, subsequently annealing at ~500 K for 3 min. The tunnelling spectra were obtained through a lock-in detection of the tunnelling current with a modulation voltage of 983 Hz. The conventional W tips and spin-polarized tips had been characterised on standard samples of Ag(111) and antiferromagnetic monolayer $CrTe_2$, respectively, prior to the measurements. The topographic images and STS were processed by WSxM and Igor.

**DFT calculations.** The calculations were performed using the generalized gradient approximation (GGA) for exchange-correlation potential, the projected augmented wave (PAW) method [44-45], and a plane-wave basis set as implemented in the Vienna ab initio simulation package (VASP) [46-47]. A kinetic energy cutoff of 700 (600) eV for the plane waves was used for structural optimization (calculations on the relative energies). An orthorhombic 9 × √3 supercell was used to model the spiral order with the propagation vector of Q = (0.22, 0, 0). A vacuum layer over 20 Å along the z direction was adopted to eliminate interactions among image layers. The on-site Coulomb interaction was considered with a U value of 4.2 eV and a J value of 0.8 eV for Ni 3d orbitals, according to the literature [48-49] and our test calculations [39]. Grimme's semiempirical D3 scheme [50] for dispersion correction was employed in combination with the Perdew–Burke–Ernzerhof functional (PBE-D3). Spin-orbit coupling (SOC) was considered in all calculations. All atoms, lattice volumes, and shapes in each supercell were allowed to relax until the residual force on each atom was less than 0.01 eV/Å.

**References:**

[1]   Mermin, N. D. & Wagner, H. Absence of ferromagnetism or antiferromagnetism in one- or two-dimensional isotropic Heisenberg models. *Phys. Rev. Lett.* **17**, 1133 (1966).

[2]   Huang, B. *et al.* Layer-dependent ferromagnetism in a van der Waals crystal down to the monolayer limit. *Nature* **546**, 270-273 (2017).

[3]   Gong, C. *et al.* Discovery of intrinsic ferromagnetism in two-dimensional van der Waals crystals. *Nature* **546**, 265-269 (2017).

[4]   Gong, C. & Zhang, X. Two-dimensional magnetic crystals and emergent heterostructure devices. *Science* **363**, 706 (2019).

[5]  Burch, K. S., Mandrus, D. & Park, J. G. Magnetism in two-dimensional van der Waals materials. *Nature* **563**, 47-52 (2018).

[6]  Chen, W.J. *et al.* Direct observation of van der Waals stacking–dependent interlayer magnetism. *Science* **366**, 983 (2019).

[7]  Xian, J. J. *et al.* Spin mapping of intralayer antiferromagnetism and field-induced spin reorientation in monolayer $CrTe_2$. *Nat. Commun.* **13**, 257 (2022).

[8]  Deng, Y. J. *et al.* Gate-tunable room-temperature ferromagnetism in two-dimensional $Fe_3GeTe_2$. *Nature* **563**, 94-99 (2018).

[9]  Tang, M. *et al.* Continuous manipulation of magnetic anisotropy in a van der Waals ferromagnet via electrical gating. *Nat. Electron.* **6**, 28-36 (2023).

[10] Zhang, W. *et al.* Spin-Resolved Imaging of Antiferromagnetic Order in $Fe_4Se_5$ Ultrathin Films on $SrTiO_3$. *Adv. Mater.* **35**, 2209931 (2023).

[11] Chua, R. *et al.* Can Reconstructed Se-Deficient Line Defects in Monolayer $VSe_2$ Induce Magnetism? *Adv. Mater.* **32**, 2000693 (2020).

[12] Bikaljević, D. *et al.* Noncollinear Magnetic Order in Two-Dimensional $NiBr_2$ Films Grown on Au(111). *ACS Nano* **15**, 14985-14995 (2021)

[13] Xie, H. *et al*. Evidence of non-collinear spin texture in magnetic moiré superlattices. *Nat. Phys*. **19**, 1150-1155 (2023).

[14] Song, Q. *et al.* Evidence for a single-layer van der Waals multiferroic. *Nature* **602**, 601 (2022).

[15] Von Bergmann, K. *et al.* Interface-induced chiral domain walls, spin spirals and skyrmions revealed by spin-polarized scanning tunneling microscopy. *J. Phys.:*

*Condens. Matter.* **26**, 394002 (2014)

[16]    Wulfhekel, W. & Gao, C. L. Investigation of non-collinear spin states with scanning tunneling microscopy. *J. Phys.: Condens. Matter.* **22**, 084021 (2010)

[17]    Song, T. *et al.* Direct visualization of magnetic domains and moiré magnetism in twisted 2D magnets. *Science* **374**, 1140-1144 (2021).

[18]    Hejazi, K., Luo, Z. X. & Balents, L. Noncollinear phases in moiré magnets. *Proc. Natl. Acad. Sci. U.S.A*. **117**, 10721-10726 (2020).

[19]    Wiesendanger, R. Spin mapping at the nanoscale and atomic scale. *Rev. Mod. Phys*. **81**, 1495 (2009).

[20]    Kurumaji, T. Spiral spin structures and skyrmions in multiferroics. *Multiferroics: Fundamentals and Applications* **89**, 126 (2021).

[21]    Kuindersma, S., Sanchez, J. & Haas, C. Magnetic and structural investigations on $NiI_2$ and $CoI_2$. *Physica B+C* **111** (2-3), 231-248 (1981).

[22]    Billerey D. *et al.* Neutron diffraction study and specific heat of antiferromagnetic $NiI_2$. *Phys. Rev. A* **61**, 138-140 (1977).

[23]    Billerey D. *et al.* Magnetic phase transition in anhydrous $NiI_2$. *Phys. Rev. A* **77**, 59-60 (1980).

[24]    Wu, S. *et al.* Layer thickness crossover of type-II multiferroic magnetism in $NiI_2$. *arXiv*: 2307.10686.

[25]    Botana, A. S. & Norman, M. R. Electronic structure and magnetism of transition metal dihalides: bulk to monolayer. *Phys. Rev. Materials* **3**, 044001 (2019).

[26]    Lu, M. *et al.* Mechanical, Electronic, and Magnetic Properties of $NiX_2$ (X = Cl,

Br, I) Layers. *ACS Omega* **4**, 5714 (2019).

[27] Kulish, V. V. & Huang, W. Single-layer metal halides $MX_2$ (X = Cl, Br, I): stability and tunable magnetism from first principles and Monte Carlo simulations. *Mater. Chem. C* **5**, 8734 (2017).

[28] Ni, J. Y. *et al.* Giant Biquadratic Exchange in 2D Magnets and Its Role in Stabilizing Ferromagnetism of NiCl2 Monolayers. *Phys. Rev. Lett.* **127**, 247204 (2021).

[29] Li, X. *et al.* Realistic Spin Model for Multiferroic $NiI_2$. *Phys. Rev. Lett.* **131**, 036701 (2023).

[30] Fumega, A. O. & Lado, J. L. Microscopic origin of multiferroic order in monolayer $NiI_2$. *2D Mater*. **9** 025010 (2022).

[31] Wyckoff, R.W.G. *Crystal Structures* John Wiley, New York, (1963).

[32] Cai, M. *et al.* Manipulating single excess electrons in monolayer transition metal dihalide. *Nat. Commun.* **14**, 3691 (2023).

[33] Wiesendanger, R. Spin mapping at the nanoscale and atomic scale. *Rev. Mod. Phys*. **81**, 1495 (2009).

[34] Kubetzka A. *et al.* Revealing antiferromagnetic order of the Fe monolayer on W(001): Spin-polarized scanning tunneling microscopy and first-principles calculations. *Phys. Rev. Lett.* **94**, 87204-87204 (2005).

[35] Romming N. *et al.* Writing and Deleting Single Magnetic Skyrmions. *Science* **341** 6146, 636-639 (2013).

[36] Heinze, S. *et al.* Spontaneous atomic-scale magnetic skyrmion lattice in two

dimensions. *Nature Phys* **7**, 713–718 (2011).

[37] Wang W. H. *et al.* A Centrosymmetric Hexagonal Magnet with Superstable Biskyrmion Magnetic Nanodomains in a Wide Temperature Range of 100–340 K. *Adv. Mater.* **28**, 6887 (2016).

[38] Yu X. Z. *et al.* Real-space observation of a two-dimensional skyrmion crystal. *Nature* **65**, 901 (2010)

[39] Liu, N. S. *et al.* Competing Multiferroic Phases in NiI$_2$ Mono- and Few-layers. *arXiv*: 2211.14423.

[40] Okubo, T., Chung, S. & Kawamura, H. Multiple-q States and the Skyrmion Lattice of the Triangular-Lattice Heisenberg Antiferromagnet under Magnetic Fields. *Phys. Rev. Lett.* **108**, 017206 (2012).

[41] Zhang, Z. M. *et al.* Atomic visualization and switching of ferroelectric order in β-In2Se3 films at the single layer limit. *Adv. Mater.* **34**, 2106951 (2022).

[42] Chang, K. *et al.* Discovery of robust in-plane ferroelectricity in atomic-thick SnTe. *Science* **353**, 274 (2016).

[43] Yang, X. *et al.* Possible phason-ploarons in purely one-dimensional charge order of Mo6Se6 nanowires. *Phys. Rev. X* **10**, 031061 (2020).

[44] Blöchl, P. E. Projector augmented-wave method. *Phys. Rev. B* **50**, 17953 (1994).

[45] Kresse, G. & Joubert, D. From ultrasoft pseudopotentials to the projector augmented-wave method. *Phys. Rev. B* **59**, 1758 (1999).

[46] Kresse, G. & Furthmüller, J. Efficient iterative schemes for ab initio total-energy calculations using a plane-wave basis set. *Phys. Rev. B* **54**, 11169 (1996).

[47]   Kresse, G. & Furthmüller, J. Efficiency of ab-initio total energy calculations for metals and semiconductors using a plane-wave basis set. *Comp. Mater. Sci.* **6**, 15 (1996).

[48]   Solovyev, I. V., Dederichs, P. H. & Anisimov, V. I. Corrected atomic limit in the local-density approximation and the electronic structure of d impurities in Rb. *Phys. Rev. B* **50**, 16861 (1994).

[49]   Botana, A. S. & Norman, M. R. Electronic structure and magnetism of transition metal dihalides: Bulk to monolayer. *Phys. Rev. Mater.* **3**, 044001 (2019).

[50]   Grimme, S. Semiempirical GGA-type density functional constructed with a long-range dispersion correction. *Comput. Chem.* **27**, 1787 (2006).